\documentclass[twocolumn,showpacs,aps]{revtex4}
\usepackage{graphicx}
\begin{document}
\title{Unicyclic Components in Random Graphs} 
\author{E.~Ben-Naim}
\affiliation{Theoretical Division and Center for Nonlinear Studies,
Los Alamos National Laboratory, Los Alamos, New Mexico 87545, USA}
\author{P.~L.~Krapivsky} 
\affiliation{Center for Polymer Studies and Department of Physics,
Boston University, Boston, Massachusetts 02215, USA}
\begin{abstract}
  The distribution of unicyclic components in a random graph is
  obtained analytically. The number of unicyclic components of a given
  size approaches a self-similar form in the vicinity of the gelation
  transition.  At the gelation point, this distribution decays
  algebraically, $U_k\simeq (4k)^{-1}$ for $k\gg 1$. As a result, the
  total number of unicyclic components grows logarithmically with the
  system size.
\end{abstract}


\maketitle

Random graphs underly processes such as polymerization \cite{pjf},
percolation \cite{ds}, and the formation of social networks
\cite{sp,gn}.  Random graphs have been extensively studied, especially
in theoretical computer science \cite{bb,jlr}. Special families of
random graphs have been also examined, e.g., planar random graphs
appear in combinatorics \cite{T,CS} and in physics \cite{Amb}.  The
basic framework for generic random graphs naturally emerged in two
different contexts \cite{da}. Flory \cite{pjf1,pjf} and Stockmayer
\cite{whs} modeled a polymerization process in which monomers
polymerize via binary chemical reactions until a giant polymer
network, namely a gel, emerges. Erd\H os and R\'enyi studied an
equivalent process in which connected components emerge from ensembles
of nodes that are linked sequentially and randomly in pairs \cite{er}.

Different methodologies have been employed to characterize random
graphs. Kinetic theory, specifically, the rate equation approach, was
used to obtain the size distribution of components \cite{jbm}. Using
probability theory, a number of additional characteristics including
in particular the complexity of random graphs have been addressed
\cite{bb,jlr}.

In this study, we show that the rate equation approach is useful for
studying the complexity of random graphs. Our main result asserts that
$U_k$, the average number of unicyclic components of size $k$ in a
random evolving graph, is given by
\begin{equation}
\label{main}
U_k(t)=\frac{1}{2}\,t^k\,e^{-kt}\,\sum_{n=0}^{k-1}\frac{k^{n-1}}{n!}\,.
\end{equation}
The unicyclic components size distribution becomes self-similar as the
gelation transition is approached. At the gelation point ($t_g=1$),
the size distribution develops an algebraic large-size tail:
$U_k(1)\simeq (4k)^{-1}$.  This implies that at the gelation point, 
the total number of unicyclic components depends logarithmically on
the system size.

The random graph evolves from $N$ disconnected nodes as follows. At
each step, two nodes are selected at random and a link is drawn
between them. This linking process is repeated ad infinitum, leading
to an ensemble of components, defined as maximally connected sets of
nodes. We consider multi-graphs where the two selected nodes need not
be different so that self-connections are allowed.

\begin{figure}[t]
\includegraphics*[width=4.5cm]{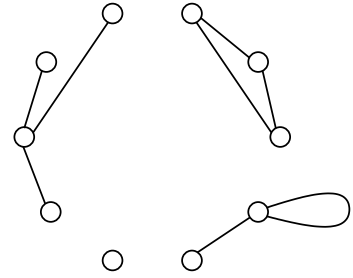}
\caption{A random graph with $N=10$ nodes and $8$ links. The two
components on the left are trees, the two components on the right are
unicyclic.}
\label{graph}
\end{figure}

Each component has a certain number of nodes and links.  The number of
nodes is the component size. The number of links minus the number of
nodes is the component complexity \cite{jan}. Up to a sign, the
complexity is the Euler characteristic of the component --- a
topological invariant.  Trees have complexity $-1$, unicyclic
components have complexity $0$, bi-cyclic components have complexity
$1$, etc (Fig.~\ref{graph}).

This linking process is treated dynamically.  Initially, there are $N$
components of size $1$ and complexity $-1$. Links are drawn between
any two nodes with a uniform rate, set equal to $1/(2N)$ without loss
of generality. It is useful to consider linking between different
components and linking within the same component separately.  Two
different components of size and complexity $(i,l)$ and $(j,m)$
respectively merge with rate $ij/(2N)$, symbolically represented by
the reaction scheme
\begin{equation}
\label{merge} 
(i,l)+(j,m)\to (i+j,l+m+1).
\end{equation}
When linking involves two nodes in the same $(k,n)$ component, the
process is
\begin{equation}
\label{inlink} 
(k,n)\to (k,n+1)
\end{equation}
and it occurs with rate $k^2/(2N)$. The average number of links at
time $t$ is $Nt/2$; the average number of self-links is smaller by a
factor $N$, i.e.  it is equal to $t/2$.

Let the number of components of size $k$ and complexity $n$ be
$N_{k,n}$. Following the dynamical rules
(\ref{merge})--(\ref{inlink}), the quantities $N_{k,n}$ change with
time according to the bi-variate Smoluchowsky equation \cite{mvs,kb}
\begin{eqnarray}
\label{biagg}
\frac{dN_{k,n}}{dt}&=&\frac{1}{2N}\,k^2(N_{k,n-1}-N_{k,n})\\
&+&\frac{1}{2N}\sum_{i+j=k}\sum_{l+m=n-1}ijN_{i,l}N_{j,m}\nonumber
-kN_{k,n}.
\end{eqnarray}
The initial condition is $N_{k,n}=N\delta_{k,0}\delta_{n,-1}$.  This
rate equation implies that the number of trees, unicyclic components,
and bi-cyclic components are proportional to $N$, $N^0$, and $N^{-1}$,
respectively.  The random graph consists primarily of trees and
unicyclic components, with more complex components being rare. Our
goal is to derive the distribution of unicyclic components. Given the
recursive nature of the rate equations, it requires the size
distribution of trees, that to leading order in $N$ equals the total
size distribution.

We first recapitulate the computation of the total size distribution.
Let $c_k=N_k/N$ be the size distribution of all components (formally,
$N_k=\sum_l N_{k,l}$). Since linking within a component affects its
complexity but does not affect its size, the corresponding (first two)
terms in the rate equation (\ref{biagg}) are irrelevant and the total
size distribution satisfies the {\it nonlinear} equation 
\begin{equation}
\label{ckt-eq-rg}
\frac{dc_k}{dt}=\frac{1}{2}\sum_{i+j=k}ijc_ic_j-k\,c_k.
\end{equation}
The initial condition is $c_k(0)=\delta_{k,1}$. The size distribution
is obtained via the generating function
\begin{equation}
\label{Fzt}
F(z,t)=\sum_{k\geq 1} kc_k(t)\,e^{kz}.
\end{equation}
This generating function evolves according to
\begin{equation}
\label{Fzt-eq}
\frac{\partial F}{\partial t}=(F-1)\, \frac{\partial F}{\partial z}\,,
\end{equation}
subject to the initial condition $F(z,0)=e^z$.  Let us re-write the
derivatives through Jacobians: $\frac{\partial F}{\partial t}=
\frac{\partial (F, z)}{\partial (t, z)}$ and $\frac{\partial
F}{\partial z}= \frac{\partial (F, t)}{\partial (z, t)}$.  Using the
relation $\frac{\partial z}{\partial t}= \frac{\partial (z,
F)}{\partial (t, F)}$, we recast the nonlinear equation (\ref{Fzt-eq})
for $F(z,t)$ into a linear equation $\frac{\partial z}{\partial
t}=1-F$ for $z(F,t)$.  Integration over time yields $z(F,t)=(1-F)t+\ln
F$.  (The integration constant $\ln F$ follows from the initial
condition $F(z,0)=e^{z}$.)  Exponentiating this equality yields an
implicit relation satisfied by the generating function
\begin{equation}
\label{fzt}
F\,e^{-tF}=e^{-t}\,\zeta, \qquad \zeta=e^z.
\end{equation}
We can expand $\zeta$ in terms of $F$, yet we are seeking the
opposite: $F=\sum kc_k(t)\zeta^k$. The size distribution can be
obtained either using the Lagrange inversion formula \cite{wilf} or
alternatively, by performing the direct calculation
\begin{eqnarray*}
kc_k&=&\frac{1}{2\pi\,i}\oint d\zeta\,\frac{F}{\zeta^{k+1}}\\
    &=&\frac{1}{2\pi\,i}\oint dF\,\frac{\zeta'(F)\,F}{[\zeta(F)]^{k+1}}\\
    &=&\frac{1}{2\pi\,i}\oint dF\,\frac{(1-tF)\,e^{ktF}}{F^k\, e^{kt}}\\
    &=&e^{-kt}\left[\frac{(kt)^{k-1}}{(k-1)!}-t\,\frac{(kt)^{k-2}}{(k-2)!}\right].
\end{eqnarray*}
The size distribution is therefore \cite{jbm,hez}
\begin{equation}
\label{ckt-rg}
c_k(t)=\frac{(kt)^{k-1}}{k\cdot k!}\,e^{-kt}.
\end{equation}

At time $t_g=1$, the system undergoes a gelation transition: a giant
component that eventually engulfs the entire mass in the system
emerges.  Close to the gelation time, the size distribution attains
the scaling behavior $c_k(t)\simeq k_*^{-5/2}\Phi(k/k_*)$ with the
typical size scale $k_*=(1-t)^{-2}$. This size scale diverges as the
gelation point is approached. The underlying scaling function
$\Phi(z)=(2\pi)^{-1/2} z^{-5/2}e^{-z/2}$ exhibits an exponential
large-size decay in the pre-gel regime while at the gelation time it
develops an algebraic tail: $c_k\sim k^{-5/2}$. This power-law
behavior allows us to estimate the size of the giant component, 
$N_g\sim N^{2/3}$ \cite{bb}, using the extremal statistics criterion
$N\sum_{k\geq N_g} c_k\sim 1$ \cite{bk}. The time when the giant
component emerges in a finite system is estimated from $N_g\sim
(1-t_g)^{-2}$, i.e., $1-t_g\sim N^{-1/3}$.

The average size distribution of unicyclic components $U_k\equiv
\langle N_{k,0}\rangle $ is coupled to the total size distribution
$c_k$. {}From (\ref{biagg}) we find that $U_k$ satisfies the {\it
linear} inhomogeneous equation
\begin{equation}
\label{Ukt}
\frac{dU_k}{dt}=\frac{1}{2}k^2c_k+\sum_{i+j=k}iU_i\,jc_j-k\,U_k. 
\end{equation}
The initial condition is $U_k(0)=0$. The first term on the right-hand side of
(\ref{Ukt}) plays the role of a source --- it represents the formation of
unicyclic components from trees via the linking of two nodes within the same
component. Such linking occurs with rate $k^2/{2N}$ (recall that multigraphs
where two nodes can be connected by more than one link are considered).  The
next two terms account for changes in the size of a unicyclic component due
to mergers with different components.

\begin{figure}[t]
\includegraphics*[width=7.5cm]{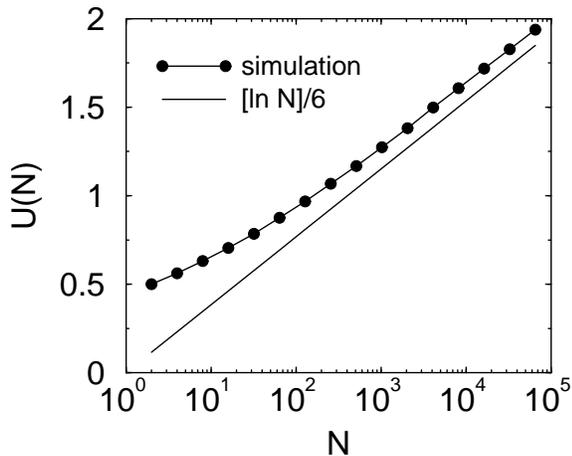}
\caption{The average number of unicyclic components at the gelation
point versus the system size $N$.}
\label{un}
\end{figure}

Consider the average number of unicyclic components $U=\sum_k U_k$
prior to the gelation time. Summing Eqs.~(\ref{Ukt}) we find that $U$
satisfies
\begin{equation}
\label{Ut}
\frac{dU}{dt}=\frac{1}{2}\,M_2
\end{equation}
where $M_2(t)=\sum_k k^2 c_k(t)$ is the second moment of the size
distribution. Using $M_2(t)=F'(z,t)|_{z=0}$ and Eq.~(\ref{fzt}) we get
$M_2=(1-t)^{-1}$. The total number of unicyclic components 
is therefore
\begin{equation}
\label{Ut-sol}
U(t)=\frac{1}{2}\ln \frac{1}{1-t}\,.
\end{equation} 
The number of unicyclic components diverges as the gelation point is
approached. The total number of unicyclic components at the gelation point is
obtained from the estimate $1-t_g\sim N^{-1/3}$; it diverges logarithmically
with the system size $N$ \cite{jklp,sj}
\begin{equation}
\label{U-gel}
U(t_g)\simeq \frac{1}{6}\,\ln N.
\end{equation}
This number is much larger compared with the average number of
self-connections that equals $1/2$ at the gelation point. The approach
to the asymptotic behavior (\ref{U-gel}) is shown using Monte Carlo
simulations (Fig.~\ref{un}). The data represent an average over $10^6$
independent realizations.

To determine the entire distribution $U_k$, it is useful to manually
solve for small $k$. The expressions 
\begin{eqnarray*}
U_1=\frac{1}{2}\,t\,e^{-t},\quad
U_2=\frac{3}{4}\,t^2\,e^{-2t},\quad
U_3=\frac{17}{12}\,t^3\,e^{-3t}
\end{eqnarray*}
suggest the following structure \hbox{$kU_k(t)=A_k
t^k\,e^{-kt}$}. Substituting this form and the size distribution
(\ref{ckt-rg}) into (\ref{Ukt}) we obtain a recurrence for the
coefficients
\begin{equation}
\label{Ak}
A_k=\frac{1}{2}\frac{k^k}{k!}+\sum_{i+j=k}A_i\,\frac{j^{j-1}}{j!}\,.
\end{equation}
To solve this recursion, we introduce the generating functions
$A(z)=\sum_{k\geq 1} A_k\,e^{kz}$ and $H(z)=\sum_{k\geq 1}
\frac{k^{k-1}}{k!}\,e^{kz}$, thereby recasting (\ref{Ak}) into
\begin{equation}
\label{AH} A=\frac{1}{2}(1-H)^{-1}\,\frac{dH}{dz}\,.
\end{equation}
The generating function $H(z)=F(z+1,1)$ is expressed via the known
generating function $F(z,t)$ and therefore,
\begin{equation}
\label{Hz}
H\,e^{-H}=\zeta, \qquad \zeta=e^z.
\end{equation}
Re-writing this equality as $\ln H - H=z$ and differentiating with
respect to $z$ we obtain $\frac{dH}{dz}=H/(1-H)$, which is then
inserted into (\ref{AH}) to give
$A=\frac{1}{2}\,H(1-H)^{-2}$. Combining this equation together with
Eq.~(\ref{Hz}), the coefficients are evaluated similar to the
derivation of the size distribution $c_k$:
\begin{eqnarray*}
A_k&=&\frac{1}{2\pi\,i}\oint d\zeta\,\,\frac{A}{\zeta^{k+1}}\\
   &=&\frac{1}{2\pi\,i}\oint dH\,\,\frac{\zeta'(H)\,A}{[\zeta(H)]^{k+1}}\\
   &=&\frac{1}{2\pi\,i}\oint dH\,\,\frac{e^{kH}}{H^k}\,\frac{1}{2(1-H)}\\
   &=&\frac{1}{2}\,\sum_{n=0}^{k-1}\frac{k^n}{n!}\,.
\end{eqnarray*}
Thus, the size distribution of the unicyclic components (\ref{main})
is obtained analytically.

Given the scaling behavior of the size distribution, we investigate
the unicyclic size distribution in the vicinity of the gelation
point. To obtain the large-$k$ behavior, the size distribution is
re-written as 
\begin{eqnarray}
U_k(t)=\frac{1}{2}t^k\,e^{-kt}\,\frac{k^{k-2}}{(k-1)!}
\sum_{j=0}^{k-1}\prod_{i=0}^j\left(1-\frac{i}{k}\right).
\end{eqnarray}
When $k\gg 1$, the product approaches an exponential, $e^{-j(j+1)/2k}$, 
and the summation can be replaced by integration
\begin{eqnarray*}
\sum_{j=0}^{k-1}\prod_{i=0}^j\left(1-\frac{i}{k}\right)
&\simeq& \int_0^k dj\,\,\exp\left[-\frac{j(j+1)}{2k}\right]\\
&\simeq& \int_0^\infty dj\,\,\exp\left[-\frac{j^2}{2k}\right]=\sqrt{\pi k/2}\,.
\end{eqnarray*}

The leading large-$k$ asymptotics of the unicyclic components
size distribution is therefore
\begin{equation}
\label{Ukt-large}
U_k(t)\simeq \sqrt{\frac{\pi}{8k}}\,\,\frac{(kt)^k}{k!}\,e^{-kt}\,.
\end{equation}
In the vicinity of the gelation point, this distribution approaches
the scaling form $U_k(t)\simeq k_*^{-1}\Psi(k/k_*)$ with the same
scaling variable as the one underlying the total size distribution, $k_*=
(1-t)^{-2}$. The scaling function is also similar in form
\begin{equation}
\label{Psi}
\Psi(z)=(4z)^{-1}\,e^{-z/2}.
\end{equation}
The unicyclic size distribution has exponential tails both above and
below the gelation transition and it becomes algebraic at the gelation
point
\begin{equation}
\label{Uk-gel}
U_k(t_g)\simeq (4k)^{-1}\qquad {\rm when}\qquad k\gg 1.
\end{equation}
This result is of course consistent with the total number of unicyclic
components: $U(t_g)\simeq\sum_{k\leq N^{2/3}}U_k(t_g)$ leads to
Eq.~(\ref{U-gel}).

We now probe the complexity of the giant component. In the
post-gelation region the gel mass (mass not contained in finite
components) $g(t)$ is given by $g=1-\sum_k
k\,c_k(t)=1-F(z,t)|_{z=0}$. Using Eq.~(\ref{fzt}), the gel mass obeys
$g=1-e^{-gt}$ for $t>1$. The complexity of the giant component
increases due to linking processes involving its internal nodes. The
total number of nodes in the giant component is $Ng$. The linking rate
is therefore $(2N)^{-1}\times (Ng)^2$, so the complexity of the giant
component is
\begin{equation}
\label{complexity}
C(t)=\frac{N}{2}\int_1^t dt'\,g^2(t')\,.
\end{equation}
Just above the gelation transition, the gel mass increases linearly,
$g\simeq 2(t-1)$, and therefore, \hbox{$C(t)\simeq
\frac{2}{3}\,N(t-1)^3$} as $t\downarrow 1$. In the large time limit
there are a few monomers apart from the giant component. The total
number of links is $Nt/2$ and the total number of nodes is
$N-N_1=N(1-c_1)$, so the complexity of the giant component grows
according to $N(t/2-1+e^{-t})$ when $t\to\infty$. This result also
follows from the general formula (\ref{complexity}) and the asymptotic
behavior of the gel mass $g(t)$.

In this study, we considered size and complexity characteristics
obtained via an average over infinitely many realizations of the
linking process. For some quantities, averaging is irrelevant in the
thermodynamic limit $N\to\infty$. For instance, the tree distribution
is an extensive random quantity ($\langle N_{k,-1}\rangle=Nc_k$) with
fluctuations of the order of $N^{1/2}$ \cite{aal,ve}, so relative
fluctuations decrease with the systems size according to
$N^{-1/2}$. Thus, for trees the average distribution $Nc_k$ well
represents the outcome of a single realization of the random evolving
graph.  For unicyclic components, fluctuations are of the same order
as the average and an analytical computation of the correlation
function $\langle N_{i,0}(t) N_{j,0}(t)\rangle$ is a challenging open
problem. This correlation function is required for the determination
of the average distribution of bi-cyclic components, so the naive form
of the rate equation (\ref{biagg}) is inadequate for describing
bi-cyclic (and more complex) components.

In conclusion, we have obtained the size distribution of unicyclic
components in a random graph. Overall, the unicyclic size distribution
has similar properties as the total size distribution; It becomes
self-similar near the gelation point, and generally, it has an
exponential tail. Precisely at the gelation point the distribution has
an algebraic tail. The main difference with the size distribution is
in the value of the power-law exponent itself.

This research was supported in part by DOE (W-7405-ENG-36).

\end{document}